\documentclass[preprint,amsmath,amssymb]{revtex4}

\usepackage{latexsym} 
\usepackage{amssymb}  
\usepackage{amsbsy}   
\usepackage{graphicx}       
\usepackage{multirow}  
\usepackage[dvips]{color}
\usepackage{bm}   
\usepackage{color}
\usepackage{subfigure}

\newcommand{\al}{\alpha}
\newcommand{\be}{\beta}
\newcommand{\ga}{\gamma}
\newcommand{\Ga}{\Gamma}
\newcommand{\de}{\delta}
\newcommand{\De}{\Delta}

\newcommand{\eps}{\epsilon}

\newcommand{\la}{\lambda}

\newcommand{\Si}{\Sigma}
\renewcommand{\th}{\theta}   

\newcommand{\om}{\omega}

\newcommand{\beq}{\begin{equation}}
\newcommand{\eeq}{\end{equation}}
\newcommand{\ba}{\begin{array}}
\newcommand{\ea}{\end{array}}
\newcommand{\bea}{\begin{eqnarray}}
\newcommand{\eea}{\end{eqnarray}}
\newcommand{\bi}{\begin{itemize}}  
\newcommand{\ei}{\end{itemize}}
\newcommand{\ben}{\begin{enumerate}} 
\newcommand{\een}{\end{enumerate}}
\newcommand{\bc}{\begin{center}}
\newcommand{\ec}{\end{center}}

\newcommand{\p}{\partial}
 
\renewcommand{\>}{\rangle} 
\newcommand{\txt}{\textstyle}
\newcommand{\dsp}{\displaystyle}

\newcommand\eqn[1]{(\ref{#1})}      
\newcommand\Eqn[1]{Eq.~(\ref{#1})}  

\newcommand{\half} {{\txt \frac{1}{2}}}

\newcommand{\tr}{\mbox{tr}}
\newcommand{\Tr}{\mbox{Tr}}

\newcommand{\nn}{\nonumber \\}
\newcommand{\ee}[1]{\times 10^{#1}}

\newcommand{\km}{{\rm km}}

\newcommand{\MeV}{{\rm MeV}}

\newcommand{\dm}{\delta m}
\newcommand{\Mscat}{{\cal M}}

\newcommand{\lshear}{l^{\rm shear}}
\newcommand{\lscat}{l^{\rm scat}}

\usepackage[normalem]{ulem}  

\begin{document}

 
\title{Shear viscosity from kaon condensation
in color-flavor-locked quark matter}

\author{
Mark G.~Alford$^1$, Matt Braby$^2$, Simin Mahmoodifar$^1$\\
$^1$Department of Physics, Washington University in St.~Louis, MO~63130, USA \\
$^2$Department of Physics, North Carolina State University, Raleigh, NC 27695, USA \\[2ex]
}

\date{6 Jan 2010} 

\begin{titlepage}
\renewcommand{\thepage}{}          

\begin{abstract}
We calculate the kaonic contribution to the shear viscosity of 
quark matter in the
kaon-condensed color-flavor locked phase (CFL-K0).
This contribution comes from a light pseudo-Goldstone boson which
arises from the spontaneous breaking of the flavor symmetry  by
the kaon condensate.
The other contribution, from the exactly massless superfluid ``phonon'',
has been calculated previously.
We specialize to a particular form of the interaction lagrangian,
parameterized by a single coupling. 
We find that if we make reasonable guesses for the values of the
parameters of the effective theory, the kaons have a much smaller
shear viscosity than the superfluid phonons, but also a much shorter mean free
path, so they could easily provide the dominant contribution 
to the shear viscosity of CFL-K0 quark matter in a neutron star
in the temperature range $0.01$ to $1$~MeV ($10^8$ to $10^{10}$\,K).

\end{abstract}
\maketitle
\end{titlepage}

\section{introduction}
\label{intro}

In this paper, we explore the shear viscosity of
one of the predicted phases of high density quark matter.  
Transport properties of quark matter, such as the shear viscosity,
are of interest because they are the basis for
signatures by which we could infer or rule out 
the presence of exotic phases in the core of neutron stars.
Previous work on transport properties
has studied the bulk viscosity 
\cite{Alford:2007rw,Alford:2008pb,Manuel:2007pz,Mannarelli:2009ia}, 
specific heat \cite{Alford:2007qa,Alford:2004zr},  
neutrino emissivity 
\cite{Carter:2000xf,Jaikumar:2002vg,Jaikumar:2005hy,Schmitt:2005wg,Wang:2006tg},
and thermal conductivity 
\cite{Pethick:1991mk,Shovkovy:2002kv,Horowitz:2008jt,Aguilera:2008ed,Braby:2009dw}.

The shear viscosity is phenomenologically relevant because it damps
physically important excitation modes of the star. In particular,
a fast-rotating neutron star will spin down rapidly if the 
internal viscosity is too low, because of the spontaneous growth
of r-modes \cite{Lindblom:1998wf}.  The observation of rapidly rotating
neutron stars can therefore be used to place limits on the internal 
viscosity. Calculating the viscosity of candidate phases then allows 
us say whether these phases can be present in the neutron star.

This paper studies the kaon-condensed color-flavor locked phase (CFL-K0) 
of quark matter. This is a candidate phase of quark matter at the central
density of a neutron star. For a review, see Ref.~\cite{Alford:2007xm}.
The CFL-K0 phase has a condensate of kaons \cite{Bedaque:2001je,Kaplan:2001qk}, 
which spontaneously
breaks the flavor symmetry, producing a Goldstone boson. 
Since the flavor symmetry is also explicitly
broken by the weak interaction, this Goldstone kaon acquires
a small mass in the keV range \cite{Son:2001xd}. Our analysis
is relevant to temperatures above this value, 
where one can ignore this small mass.
We use the effective theory of the Goldstone kaon,
which was worked out in Ref.~\cite{Alford:2007qa}. The full interaction
lagrangian has three independent coupling constants, but we will
specialize to a specific ratio of their values, which makes
our results dependent on one overall kaon interaction
coupling. This enables us to make an estimate of the
expected scale of the shear viscosity in this phase.
We defer the calculation for the most general
interaction lagrangian to future work.

As well as the shear viscosity
we calculate the mean free path of the Goldstone kaons, since
the concept of shear viscosity will only be applicable to
the matter in the neutron star if there is local equilibration
on distance scales much smaller than the size of the star.

The rest of the paper is laid out as follows:  Section~\ref{effective} will
discuss the low energy effective theory of the CFL-K0 phase and the
interactions among the lowest energy excitations of the theory.  
Section~\ref{sec:shear} will show the results for the contribution of the kaons
to the shear viscosity. Section~\ref{sec:conclusions} will present some
conclusions and discuss future directions of this work.  In the appendices
we cover technical
details of the calculation of the mean free path, the treatment
of the co-linear regime of the collision integral, and the approximate
evaluation of the collision integral.

\section{Low Energy Effective Theory}
\label{effective}

\subsection{Lowest-order lagrangian}

The low energy degrees of freedom in color-flavor-locked phases of
quark matter are the massless superfluid Goldstone mode, arising from
the spontaneous breaking of baryon number, and the
light pseudo-Goldstone meson octet, arising from the spontaneous breaking
of three-flavor chiral symmetry. The contribution of the superfluid
mode to transport properties has been studied previously \cite{Manuel:2007pz,Manuel:2004iv}.
We focus on the contribution from the meson
octet, described by a meson field $\Si$ whose effective lagrangian
up to second order is \cite{Bedaque:2001je,Kaplan:2001qk}
\beq
{\cal L} = 
\frac{f_\pi^2}{4}\Tr[D_0\Si\,D_0\Si^\dag - v^2 \nabla \Si\,\nabla \Si^\dag] 
+ a \frac{f_\pi^2}{2}\det M\,\Tr[{M^{-1}}(\Si+\Si^\dag)]
\label{lag}
\eeq
where $D_0\Si = \p_0 \Si - i[A,\Si]$. The Bedaque-Sh\"afer effective 
chemical potential \cite{Bedaque:2001je} is $A = -\frac{M.M}{2\mu_q}$,
$\mu_q$ is the quark chemical potential, and 
$M = {\rm diag}(m_u,m_d,m_s)$ is the quark mass matrix. At asymptotically high 
density the constants $f_\pi$, $v$, and $a$ can be determined by matching the 
effective theory to perturbative QCD, \cite{Son:1999cm,Son:2000tu}
\beq
 f_\pi^2= \frac{21-8\ln2}{18}\frac{\mu_q^2}{2\pi^2}
\approx (0.21 \mu_q)^2  \qquad \qquad 
 v\equiv v_H=\frac{1}{\sqrt{3}}\qquad \qquad
a = \frac{3 \De^2}{\pi^2 f_\pi^2} \ ,
\label{asymptotia}
\eeq
where $\De$ is the fermionic energy gap at zero temperature. 
This dependence of $a$ on $\De$ is also seen in NJL models
\cite{Kleinhaus:2007ve}, so from now on we will work in terms of
$f_\pi$ and $\De$, assuming that $a$ is given by \eqn{asymptotia}.
The meson field $\Si$ can be parameterized in terms of fields $\th_a$,
\beq
\Si = \exp(i\th/f_\pi) \ ,
\eeq
where $\th = \th_a T_a$, and $T_a$ are the Gell-Mann matrices of SU(3)
with normalization $\tr (T_a T_b) = 2 \de_{ab}$.
The $K^0$ and $K^+$ are 
the lightest mesonic degrees of freedom \cite{Son:1999cm,Son:2000tu}, and
electric neutrality disfavors the presence of charged kaons (since they
must be balanced by electrons), so we focus on the neutral 
kaons, $K^0$ and $\bar{K}^0$, corresponding to $\th_6$ and
$\th_7$. 
The zero-temperature neutral kaon mass and chemical potential 
can be deduced from the Lagrangian
\beq
\ba{rcl}
m_K^2 &=& am_u(m_d+m_s)  \ , \\
\mu_K &=& \dsp \frac{m_s^2 - m_d^2}{2\mu_q} \ .
\ea
\label{muK-def}
\eeq 
We will assume that $\mu_K>m_K$, so there is kaon condensation.
The critical temperature for kaon condensation is 
expected to be of order
tens of MeV, well above typical neutron star temperatures, and of the same
order as the critical temperature for the CFL condensate itself
\cite{Alford:2007qa}.
We will also assume, following Ref.~\cite{Alford:2007qa}, that the
condensate is small, so $\mu_K$ is only a little larger than $m_K$.
Given the uncertainty in the effective theory couplings, this is
a perfectly conceivable scenario: taking for example $\mu=400\,\MeV$,
$m_s=100\,\MeV$, $\De=80\,\MeV$ and $m_u=5\,\MeV$ (and ignoring
$m_d$ because it makes a negligible contribution) one obtains
$\mu_K=12.5\,\MeV$ and $m_K=11.0\,\MeV$.
It is then convenient to
define, following Ref.~\cite{Alford:2007qa}, an energy gap
\beq
\dm \equiv  m_K - \mu_K \ , \qquad \frac{|\dm|}{m_K}\ll 1 \ .
\eeq
Note that $\dm$ is negative in the CFL-K0 phase. 
Because $|\dm|\ll 1$
we can usually treat $\mu_K$ and $m_K$ as being identical
to leading order in $\dm$ (an exception is discussed in 
Sec.~\ref{sec:results}).

A self-consistent calculation~\cite{Alford:2007qa}
(see also \cite{Reddy:2003ap})
then yields the excitation energies in the neutral kaon sector,
\beq
E_\pm^2 = E_p^2 + \mu_K^2 \mp \sqrt{4\mu_K^2 E_p^2 + \de M^4} \ ,
\label{energy}
\eeq
where 
\beq
E_p^2 = v^2p^2 + {\bar M}^2 \ .
\eeq
In that self-consistent calculation, ${\bar M}$ and $\de M$ were
thermal masses that depended on temperature and the underlying
mass and chemical potential (see Eq.~(81) in Ref.~\cite{Alford:2007qa}).  
In this paper we are interested in low temperature range applicable
for compact stars.  In this case, the thermal masses become independent
of temperature and are given by
\bea
{\bar M}^2 &=& 2\mu_K^2 - m_K^2 \approx m_K^2\ , \nn
\de M^2 &=& \mu_K^2 - m_K^2 \approx 2 m_K |\dm|\ .
\label{2PI}
\eea
The mode with energy $E_+$ is massless: this is the Goldstone kaon.
We can define a corresponding field $\psi$ using the parameterization
\beq
\ba{rcl}
\th_6(x) &=& \bigl(\phi + \rho(x)\bigr) \cos \vartheta(x) \ , \\
\th_7(x) &=&  \bigl(\phi + \rho(x)\bigr)\sin \vartheta(x) \ , \\
\psi(x) &=& f_\pi \sin(\phi/f_\pi)\, \vartheta(x)  \ ,
\ea
\label{psidef}
\eeq
so $\phi$ is the kaon condensate, $\rho$ is the massive
radial mode, and $\vartheta$ is the angular Goldstone mode
which we have then rescaled to make a scalar field $\psi$ with a 
canonically normalized quadratic derivative term and the
conventional energy dimension of 1.
The mass of the radial modes is given by the value of $E_-(p=0) = \sqrt{6\mu_K^2-2m_K^2}
\approx 2m_K$, which is typically on the order of a few $\MeV$, so
at the 10 to 100 keV energy scale, which is relevant to neutron stars, 
it is heavily suppressed and can be ignored. The magnitude of the
kaon condensate is \cite{Alford:2007qa}
\beq
\phi^2 = 2 f_\pi^2\Bigl(1-\frac{m_K^2}{\mu_K^2}\Bigr)
 \approx 4 f_\pi^2 \frac{|\dm|}{m_K}
\label{Kcond}
\eeq

We can then linearize \Eqn{energy} to obtain
a linear dispersion relation for the Goldstone kaon,
\beq
\ba{rcl}
E(p) &=& \dsp {\nu} p \\[2ex]
\nu &\equiv&\dsp v\sqrt{\frac{{\bar M}^2 - \mu_K^2}{{\bar M}^2 + \mu_K^2}} 
 \approx  v \sqrt{\frac{|\dm|}{m_K}} \ .
\ea
\label{dispersion}
\eeq
The error involved in approximating \Eqn{energy} by \Eqn{dispersion} is 
less than 5\% for \mbox{$p<0.6\,\sqrt{m_K |\dm|}/v$}.

We will obtain the contribution to the shear viscosity from the
Goldstone kaon. For this we need its interaction lagrangian, but
it is easy to see that \eqn{lag} does not contain any interaction
terms for the field $\psi$. This follows from the fact that $\psi$,
as a Goldstone boson, must couple via derivatives, and \eqn{lag}
only goes to second order in derivatives. We therefore need to write
down higher order derivative terms in the effective theory to obtain
interactions among the Goldstone modes.

\subsection{Interaction lagrangian for the Goldstone kaons}

We obtain higher derivative terms in $\psi$ by writing down the
leading higher derivative terms in the lagrangian for $\Si$, and
using \eqn{psidef}. We keep only terms with the symmetries of the system,
namely rotational symmetry, parity, time-reversal, and the 
$SU(3)_L \otimes SU(3)_R$ chiral flavor symmetry. 
We also discard terms that, when we substitute \eqn{psidef},
will produce interactions that all involve the $\rho$ field;
an example is three-derivative terms where $\Si$ enters four times.
The allowed terms with no more than four derivatives of $\Si$ are shown
in Table~\ref{tab:couplings} (left column).
Since the effective theory breaks down at momenta of order $\De$
(for example, scattering of Goldstone bosons at that momentum
will produce quasiquarks, which are not included in the effective theory)
we expect that the momentum expansion will be in powers of
$(1/\De)\vec\nabla$ \cite{Bedaque:2001je}. 
We therefore expect the interactions in the left column of
Table~\ref{tab:couplings} to occur in the lagrangian
with coefficients $C_i f_\pi^2/\De^2$, where the $C_i$
are dimensionless coupling constants.

Using \eqn{psidef} and dropping terms that involve the
heavy field $\rho$, these six terms reduce to the three
corresponding interaction terms for $\psi$ shown in the
right column. In each case, we find two different interaction
terms for $\Si$ reduce to the same interaction term for $\psi$.
This means that the interaction lagrangian
for $\psi$ only depends on three linear combinations of couplings.
Note that in Table~\ref{tab:couplings} 
we have defined a scaled version of the kaon condensate
expectation value \eqn{psidef},
\beq
\varphi \equiv \phi/f_\pi \approx 2\sqrt{\frac{|\dm|}{m_K}}
\label{phi}
\eeq

\begin{table}[tb]
\newcommand{\st}{\rule[-2ex]{0em}{4ex}}
\begin{tabular}{l@{\qquad}lc}
\hline
 \st 1a. 
& \( \left(\Tr[D_0\Si D_0\Si^\dag]\right)^2 \)
 \phantom{XX} 
& \multirow{2}{*}{\( \dsp \frac{4}{f_\pi^4}(\p_0\psi)^4 
   + \frac{16\mu_K \sin{\varphi}}{f_\pi^3} (\p_0\psi)^3 \)} \\
\cline{1-2}
 \st 1b.  
& \( 2\Tr[(D_0\Si D_0\Si^\dag)^2] \)
& 
\\
\hline
 \st 2a.
& \( \left(\Tr[\nabla\Si\nabla\Si^\dag]\right)^2 \) 
& \multirow{2}{*}{\(  \dsp \frac{4}{f_\pi^4}(\nabla\psi)^4 \) } \\
\cline{1-2}
 \st 2b.
& \( 2\Tr[(\nabla\Si\nabla\Si^\dag)^2] \)
& 
\\
\hline
 \st 3a.
& \( \Tr[D_0\Si D_0\Si^\dag]\Tr[\nabla\Si\nabla\Si^\dag] \)
& \multirow{2}{*}{
     \phantom{XX} 
     \( \dsp \frac{4}{f_\pi^4}(\p_0\psi)^2(\nabla\psi)^2 
    + \frac{8\mu_K \sin{\varphi}}{f_\pi^3} (\p_0\psi)(\nabla \psi)^2 \) } \\
\cline{1-2}
\st 3b.
& \( 2\Tr[D_0\Si D_0\Si^\dag \nabla\Si\nabla\Si^\dag] \) 
& 
\\
\hline
\end{tabular}
\caption{The six interaction terms at fourth order in derivatives
for the effective theory (first column), and the interaction terms
for $\psi$ that they transform to using \eqn{psidef},
when terms involving $\rho$ are dropped. In the effective lagrangian
they have coefficients of order $f_\pi^2/\De^2$.
}
\label{tab:couplings}
\end{table}

The interaction Lagrangian for $\psi$ can then be written out as
\beq
\ba{rcl}
{\cal L} &=& \dsp C_1 \frac{2}{f_\pi^2\De^2}(\p_0\psi)^4 
+ C_3 \frac{2}{f_\pi^2\De^2}(\p_0\psi)^2(\nabla\psi)^2 
+ C_2 \frac{2}{f_\pi^2\De^2}(\nabla\psi)^4 \\[2ex]
&& \dsp + C_1 \frac{8\mu_K \sin \varphi}{f_\pi\De^2}(\p_0\psi)^3 
+ C_3\frac{4\mu_K \sin \varphi}{f_\pi\De^2}(\p_0\psi)(\nabla\psi)^2 .
\ea
\label{L-GBint}
\eeq
At this point we specialize to a particular form of the interaction
lagrangian, with the following
relationship among the three coupling constants.
\beq
C \equiv C_1 = C_2 = -\half C_3.
\label{guess}
\eeq
This reduces the number of coupling constants from three to one.
The remainder of our calculation is for this special case, adopted
because it leads to a particularly simple interaction lagrangian
which is similar to that written down for the superfluid phonon
in Refs.~\cite{Manuel:2004iv,Rupak:2007vp},
\beq
{\cal L}_{int} = \frac{\la}{4f_\pi^4}(\p_\mu\psi\p^\mu\psi)^2 
+ \frac{g}{2f_\pi^2}(\p_0\psi)(\p_\mu\psi\p^\mu\psi)
\label{Lint}
\eeq
where
\beq
\la = 8C \frac{f_\pi^2}{\De^2} \qquad {\rm and }\qquad 
g = 16\sin(\varphi)C \, \frac{\mu_K f_\pi}{\De^2}
\label{lambda.g}
\eeq
We leave the analysis of the fully general interaction lagrangian
\eqn{L-GBint} for future work.


\section{Shear Viscosity}
\label{sec:shear}

The shear viscosity  $\eta$ is a
coefficient in the viscous stress tensor $\de T_{ij}$, which 
is the deviation from equilibrium of the momentum flux tensor
$T_{ij}$ for a fluid with pressure $P$ and  energy density $\eps$,
\bea
T_{ij} &=& T_{ij}^{(eq)} + \de T_{ij} \nn
T_{ij}^{(eq)} &=& (P+\eps)V_i\,V_j - P\de_{ij} \nn
\de T_{ij} &=& -\eta V_{ij} + \cdots
\label{stress}
\eea
where 
\bea
V_{ij} &=& \p_i\,V_j + \p_j\,V_i - \frac{2}{3} \de_{ij} \nabla \cdot {\bf V}
\label{V}
\eea
and the ellipsis in the equation for $\de T_{ij}$
stands for other dissipative terms arising from phenomena such as
bulk viscosity and thermal conductivity. 
${\bf V}({\bf x},t)$ is the fluid velocity at a given position and time.
We will need to make sure that we only need to keep the first order
dissipative terms in \Eqn{stress} (see discussion at the end
of \ref{sec:num.results}).
The stress-energy tensor and the viscosity can be
calculated using kinetic theory \cite{Lifshitz1981}. For a system of
identical bosonic particles with dispersion relation $E_p$,
\beq
T_{ij}({\bf x},t)
=\nu^2 \int \frac{d^3 p}{(2\pi)^3} \frac{p_i\,p_j}{E_p} f_p({\bf x},t) \ .
\label{stress-kinetic}
\eeq
where $\nu$ is the velocity of the Goldstone kaon (see Eq.~\ref{dispersion}).  
The full distribution function is given by
\beq
f_p({\bf x},t) = \frac{1}{e^{p_\mu u^\mu({\bf x},t)/T}-1}
= f^0_p + \de f_p({\bf x},t)
\label{full-f}
\eeq
where $u^\mu({\bf x},t)$ is the 4-velocity of the fluid, and
$\de f_p$ is a small departure from the equilibrium Bose-Einstein distribution
\beq
f^0_p = \frac{1}{e^{E_p/T}-1} \ .
\label{Bose-Einstein}
\eeq
For shear viscosity we are interested in deviations
from equilibrium arising from a shear flow, so we write the deviation from
equilibrium as
\bea
\de f_p({\bf x},t) &=& -\frac{\chi(p,{\bf x},t)}{T}f^0_p(1+f^0_p) \nn
\chi(p,{\bf x},t) &=& g(p)\, p_{kl}\, V_{kl}({\bf x},t) ,
\label{dfp}
\eea
where 
\beq
p_{kl} = p_k\,p_l - \frac{1}{3}\de_{kl} p^2.
\label{p}
\eeq
Substituting \eqn{dfp} into \eqn{stress-kinetic} and \eqn{stress}
we find
\beq
\de T_{ij}({\bf x},t) = -\nu^2 \int \frac{d^3 p}{(2\pi)^3} \frac{p_i\,p_j}{E_p} 
\frac{g(p)\, p_{kl}}{T}f^0_p(1+f^0_p)\, V_{kl}({\bf x},t) .
\label{deltaT}
\eeq
Using the definition of $V_{ij}$ (see Eq. \ref{V}) one can write 
$\de T_{ij}$ (Eq. \ref{stress}) in the following form 
\beq
\de T_{ij} = - \frac{\eta}{2}[\de_{ik}\de_{jl}+\de_{il}\de_{jk}
             - \frac{2}{3}\de_{ij}\de_{kl}]V_{kl} ,
\eeq 
Comparing this to \Eqn{deltaT} gives us
\beq
\frac{\eta}{2}[\de_{ik}\de_{jl}+\de_{il}\de_{jk}-\frac{2}{3}\de_{ij}\de_{kl}] = \nu^2 \int \frac{d^3 p}{(2\pi)^3} \frac{p_i\,p_j}{E_p} \frac{g(p)\, p_{kl}}{T}f^0_p(1+f^0_p) .
\eeq
Then, by contracting the tensor on the left hand side with respect to the 
pairs of indices $i,k$ and $j,l$ we can determine the shear viscosity in 
terms of the function $g(p)$,
\beq
\eta = \frac{4\,\nu^2}{15\,T}\int_p\ p^4\, f^0_p(1+f^0_p)\, g(p) ,
\label{def.eta}
\eeq
where we have adopted the notation
\beq 
\int_p = \int \frac{d^3 p}{2\,E_p\,(2\,\pi)^3} .
\eeq
Using the fact that $p^4 = \frac{3}{2}p_{ij}\,p_{ij}$ (see Eq.~\ref{p}) one 
can write an alternate form of the shear viscosity which will be used later,
\beq
\eta = \frac{2\,\nu^2}{5\, T}\int_p f^0_p(1+f^0_p)\, g(p) p_{ij} p_{ij} .
\label{def.eta1}
\eeq

To solve for the viscosity, we need to find a form for $g(p)$.  To do so, we 
can use the Boltzmann equation given in the absence of external
forces by
\beq
\frac{df_p}{dt} = \frac{\p f_p}{\p t} + {\bf V} \cdot \nabla f_p 
= C[f_p] \ .
\eeq
The left-hand side can be written as \cite{Lifshitz1981}
\beq
\frac{df_p}{dt} = \nu\frac{f^0_p}{2\,p\,T}(1+f^0_p)\,p_{ij}\,V_{ij} \ .
\label{dfdt}
\eeq
Again, this specific form assumes that we are only interested in shear flows.
Thermal gradients and bulk flows would give additional term on the right hand
side of \Eqn{dfdt}.  It is this form that also helped motivate the structure
of the ansatz in \Eqn{dfp}.

The collision operator $C[f_p]$ should contain any possible
collision terms for the kaons.  We will restrict ourselves
to the terms lowest order in the coupling constants as more
vertices are suppressed because each vertex brings in more powers 
of $1/f_\pi$ or $1/\De$ (see \eqn{Lint} and \eqn{lambda.g}).
Also, we will ignore the $1\leftrightarrow 2$ processes because for a
particle with a linear dispersion relation such processes must be
co-linear, so they do not involve momentum transfer that would contribute
to the shear viscosity.
Finally we are left with the collision operator for 2-body scattering given 
by 
\cite{Rupak:2007vp}
\beq
C_{2\leftrightarrow2} = \frac{1}{2 E_p}\int_{k,k',p'} (2\pi)^4 \de^4(P+K-P'-K')
|\Mscat|^2 D_{2\leftrightarrow2}
\eeq
Where $\Mscat$ is the $2\rightarrow2$ scattering amplitude and 
$D_{2\leftrightarrow2}$  contains the distribution functions.
\beq
D_{2\leftrightarrow2} = f_{p'}\,f_{k'} (1+f_{p}) (1+f_{k})
                         - f_p\,f_k (1+f_{p'}) (1+f_{k'}).
\eeq
We can also linearize the distributions as $D \approx D^0 + \de D$ 
using our definition of $\de f_p$ in \Eqn{dfp}.  $D^0$ would make the 
collision integral vanish by detailed balance.
One can then write 
\beq
\de D_{2\leftrightarrow2} = \frac{1}{T}f^0_p\,f^0_k (1+f^0_{p'}) (1+f^0_{k'}) 
   \left(\chi(p)+\chi(k)-\chi(p')-\chi(k')\right)
\eeq
and the collision integral as
\bea
C_{2\leftrightarrow2} &\approx& \frac{f^0_p}{2 E_p T} \int_{k,p',k'} 
(2\pi)^4 \de^4(P+K-P'-K')|\Mscat|^2 f^0_k (1+f^0_{p'})(1+ f^0_{k'})\nn
&&\left[g(p)p_{ij} + g(k)k_{ij} - g(k')k'_{ij} - g(p')p'_{ij}\right] V_{ij}\nn
&\equiv& \frac{1}{2E_p T}F_{ij}[g(p)] V_{ij}
\eea
Using \eqn{dfdt} and the Boltzmann equation, we can conclude that
\beq
\nu^2 f^0_p (1+f^0_p)\,p_{ij} = F_{ij}[g(p)]
\eeq
One can then use this equation and \eqn{def.eta1}
to get another expression for the shear viscosity in terms of collision term
\beq
\eta = \frac{2}{5 T}\int_p\, g(p) p_{ij} F_{ij}[g(p)]\
\label{def.eta2}
\eeq
The process now is to evaluate $\eta$ from \eqn{def.eta} and \eqn{def.eta2}, 
and ensure that they give the same answer.  Formally, this is equivalent to 
solving the Boltzmann equation directly.  Ensuring that the two forms are 
equal is quite non-trivial and is typically done by expanding $g(p)$ 
using an orthogonal set of functions, \cite{Resibois1977,Chen:2006iga,Dobado:2001jf}
\beq
g(p) = p^n \sum_{s=0}^{N} b_s B_s(p) \ .
\label{g-pnml}
\eeq
This expansion introduces two new parameters: $N$, the order of the
polynomial approximation; and $n$, which we call the minimum-exponent parameter,
because the lowest power of $p$ that occurs in the polynomial expansion
is $p^n$.
The correct result is obtained in the limit $N\to\infty$ for any
value of $n$. However, as we will see, the rate of
convergence with increasing $N$ is strongly dependent on
the minimum-exponent parameter $n$.

The polynomials $B_s(p)$ are defined such that the coefficient of the highest 
power $p^s$ is 1 and they obey the orthogonality condition \cite{Rupak:2007vp}
\beq
\int_p f^0_p (1+f^0_p)\,p_{ij}\,p_{ij} p^n B_r(p) B_s(p) = A_{r}^{(n)} 
\de_{rs} \ .
\label{An}
\eeq
These conditions uniquely specify the $B_s(p)$ for all $s$, 
starting with $B_0(p)=1$.
From the orthogonality condition we find
\beq
A_0^{(n)} = \frac{2}{3}\int_p f^0_p(1+f^0_p) p^{4+n} 
= \frac{T^{6+n}}{6\pi^2\nu^{7+n}}\Ga(6+n)\zeta(5+n) \ .
\eeq
Using $g(p)$ from \Eqn{g-pnml} in \Eqn{def.eta}, and using the
definition of $A_r^{(n)}$ from \Eqn{An}, we get
\beq
\eta = \frac{2\, \nu^2}{5 T} b_0 A_0^{(n)} \ .
\label{eta1}
\eeq
An alternative expression for $\eta$ follows from substituting $g(p)$ 
from \Eqn{g-pnml} into \Eqn{def.eta2},
\beq
\eta =  \sum_{s,t=0}^{N=\infty} b_s b_t M_{st} \ ,
\label{eta2}
\eeq
where
\beq
\ba{rcl}
M_{st} &=&\dsp \frac{2}{5 T} \int_{p,k,k',p'}(2\pi)^4 \de^4(P+K-P'-K')|\Mscat|^2 
f^0_p f^0_k (1+f^0_{k'})(1+f^0_{p'})\ p^n B_s(p) p_{ij} \De^t_{ij} \ , \\
&=&\dsp \frac{1}{10 T} \int_{p,k,k',p'}(2\pi)^4 \de^4(P+K-P'-K')|\Mscat|^2 
f^0_p f^0_k (1+f^0_{k'})(1+f^0_{p'})\ \De^s_{ij}\De^t_{ij}
\ea
\label{Mst}
\eeq
and
\beq
\De^t_{ij} = B_t(p) p^n p_{ij} + B_t(k) k^n k_{ij} - B_t(k') k'^n k'_{ij} - 
B_t(p') p'^n p'_{ij} \ .
\label{Delta_ij}
\eeq
The second line of \Eqn{Mst} uses the symmetry under relabeling the momenta
of the legs in the scattering diagrams ($P\rightarrow K$, etc.), and can be
used to demonstrate that the diagonal elements of $M_{st}$ are positive 
definite. As we will see below, this ensures that the shear viscosity is also 
positive.

Requiring the two forms of $\eta$ to be equal leads to a matrix equation
for all the $b_i$'s.  From that we extract $b_0$,
\beq
b_0 = \frac{2\, \nu^2}{5 T} A_0^{(n)} (M^{-1})_{00}
\eeq
where $(M^{-1})_{00}$ means the first entry in the matrix inverse of $M_{st}$.
Using this in \Eqn{eta1}, we find
\beq
\eta = \frac{4\, \nu^4}{25 T^2} (A_0^{(n)})^2 (M^{-1})_{00} \ .
\label{eta.full}
\eeq
As noted above, this expression becomes accurate in the limit
$N \to \infty$, where the matrix $M$ is of infinite size.
It is known \cite{Resibois1977,Manuel:2004iv,Rupak:2007vp} that the result for
finite $N$ rises with $N$, so for a matrix $M_N$, with finite dimension 
$N$, that obeys \eqn{Mst},
\beq
\eta \geq \frac{4\, \nu^4}{25 T^2} (A_0^{(n)})^2 (M_N^{-1})_{00}
\eeq
We will see below that this expression converges rapidly 
with $N$ for the optimal choice of the minimum-exponent 
parameter $n$. 

\begin{figure}
\includegraphics[width=0.9\textwidth]{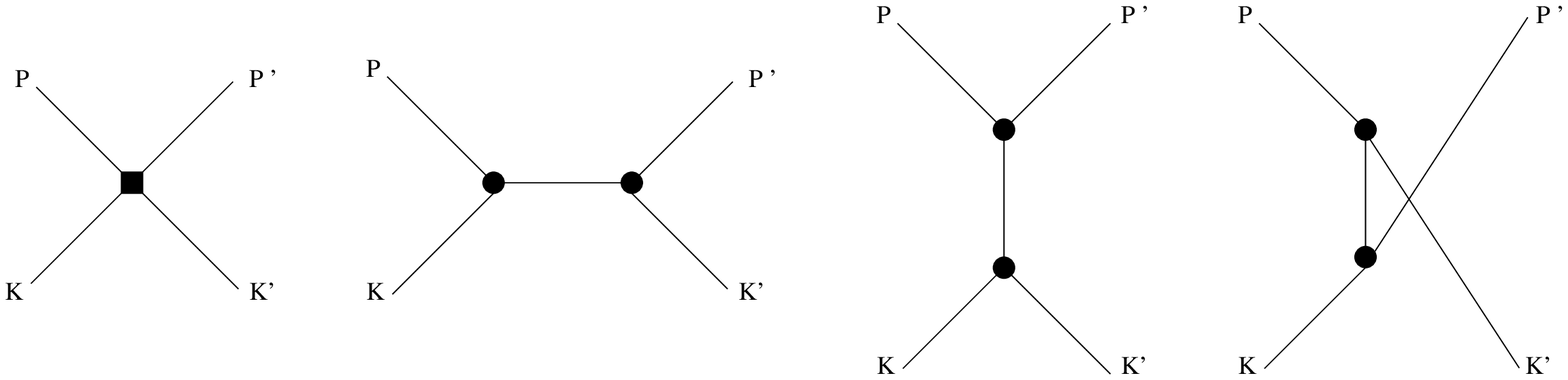}
\caption{Feynman diagrams for the 2-body scattering amplitude.  
The black square is the 4-point contact interaction, the black circles
are the 3-point vertices.
\label{fig:feyn}}
\end{figure}

The remaining task is to evaluate the integral in \Eqn{Mst}.  
This requires the matrix elements for the 
$2 \leftrightarrow 2$ scattering process, 
$i\Mscat = i\Mscat_c + i\Mscat_s
+ i\Mscat_t + i\Mscat_u$, see Fig.~\ref{fig:feyn}, with the individual channels being given by
\bea
i\Mscat_c &=&\dsp \frac{\la}{f_\pi^4}
  \left[(P\cdot K)(P'\cdot K) + (P\cdot K')(P'\cdot K) 
  + (P\cdot P')(K\cdot K')\right] \nn
i\Mscat_s &=& \dsp \frac{g^2}{f_\pi^4}
  \left[2(p_0+k_0)P\cdot K + p_0 K^2 + k_0 P^2\right]
               \left[2(p'_0+k'_0)P'\cdot K' + p'_0 K'^2 + k'_0 P'^2\right] 
                G(P+K)\nn
i\Mscat_t &=& i\Mscat_s (P\leftrightarrow K') \nn
i\Mscat_u &=& i\Mscat_t (P\leftrightarrow K), 
\label{Mscat}
\eea
where 
\beq
G(Q) = \frac{1}{(q_0^2-\nu^2q^2) + i\, {\rm Im}\, \Pi(q_0,q)}
\label{prop}
\eeq
is the Goldstone kaon propagator and the last two lines in \eqn{Mscat}
come from crossing symmetries.

The 12-dimensional integral in \Eqn{Mst} can be simplified by eliminating the 
${\bf p'}$ integral using the momentum-conserving delta-function.  Then
one can use the energy-conserving delta-function to 
eliminate the integral over the 
magnitude of $k'$. Three of the remaining 8 integrals
can be eliminated by selecting the z-axis to lie along the vector ${\bf p}$,
and noting that only the difference in the two remaining azimuthal angles 
matters.  This leaves a 5-dimensional integral over the magnitudes of $p$ and 
$k$, two polar angles corresponding to $k$ and $k'$ and one azimuthal angle.  
This can be evaluated numerically (see appendix \ref{app:coll-int})
using the {\em Vegas} Monte Carlo algorithm
\cite{PeterLepage1978192,Vegas}. 

The results that we present below are obtained by setting the
minimum-exponent parameter $n$ to $-1$. This value is expected to
give optimal convergence of the calculated shear viscosity to its physical
value as a function of $N$ because, as shown in appendix~\ref{app:integrals}, 
this term most strongly suppresses the co-linear scattering and 
therefore give the smallest collision term.  The shear viscosity is 
inversely proportional to the collision term and since we have a variational 
procedure that says the answer we get is a lower bound, we are only 
interested in the largest value of the shear viscosity that we can 
calculate. 

To check this reasoning we show in Table~\ref{tab:res}
results of calculations of the shear viscosity for different
values of $n$ and $N$.  We see that for $n=-1$ the value of $\eta$ at
low $N$ is already close to the maximum (asymptotic) value at $N=\infty$.
For $n=-2$ the convergence is almost as good, achieving 
$\sim 1\%$ accuracy
at $N=2$. For other values of $n$ the convergence is dramatically poorer.
This behavior was also seen in Refs.~\cite{Manuel:2004iv,Rupak:2007vp}.
We conclude that we can achieve accuracy of better than 1\% by choosing
$n=-1$, and only keeping the first polynomial ($N=0$),
i.e.~we set $g(p) = 1/p$.  

\begin{table}[tb]
\begin{tabular}{c@{\qquad}c@{\qquad}c@{\qquad}c}
\hline
$n$ & $N=0$ & $N=1$ & $N=2$ \\ \hline
-3 & 0.059 & 0.861 & 2.22 \\ \hline
-2 & $1.59 \times 10^5$ & $2.01 \times 10^5$ & $2.02 \times 10^5$\\ \hline
-1 & $2.04 \times 10^5$ & $2.04 \times 10^5$ & $2.04 \times 10^5$\\ \hline
0  & 0.610 & 4.70 &  19.4\\ 
\hline
\end{tabular}
\caption{Values of shear viscosity in $(\MeV)^3$ as a function of the order $N$
of the polynomial approximation to $g(p)$, for different choices of the 
minimum-exponent parameter $n$. The calculated
value rises towards the physical result 
as $N\to\infty$, and in this limit should be independent of $n$. 
We see that for $n=-1$ the result converges 
very rapidly as $N$ rises, but
for other values of $n$ the convergence is slower.
\label{tab:res}}
\end{table}


\section{Results}
\label{sec:results}

\subsection{Analytic results}
\label{sec:analytic}

We now describe how the shear viscosity depends on the 
temperature and on the parameters of the effective lagrangian for
the Goldstone kaon. The relevant parameters are the speed
$\nu$ \eqn{dispersion} of the kaon and its interaction couplings
$\la$ and $g$ \eqn{Lint}, which in turn depend on more
basic parameters $C$, $f_\pi$, $\De$, $m_K$ \eqn{lambda.g}.
Recall that we have assumed $|\dm|\ll 1$, so in the expressions below,
$\mu_K$ and $m_K$ are usually interchangeable. One exception is
the phonon speed $\nu$, which occurs in the shear viscosity
raised to the 11th power (see discussion after \eqn{eta-form})
so we use the full expression (the identity in \eqn{dispersion})
for it.

Before performing any numerical calculations, we can extract the
temperature dependence of the shear viscosity. Because the co-linear
scattering will not contribute to the answer, the propagator does not
need to be regulated by the self-energy.  Since the
temperature only appears in the distribution functions and the
self-energy, we can now factorize out the temperature dependence by 
rescaling all the momenta by the temperature.  Doing so we find
\beq
M_{st} \sim \dsp  T^{15+2\,n + s+t} \ .
\eeq
where $s$ and $t$ are the indices of the matrix indicating how
many terms we are keeping in our expansion for $g(p)$ and $n$ is the 
minimum-exponent parameter (see \eqn{g-pnml}).
We also recall the temperature dependence of $A_r^{(n)}$ from \Eqn{An},
\beq
A_0^{(n)} \sim T^{n+6} \ .
\eeq
Therefore, from \Eqn{eta.full}, we obtain the temperature dependence of the 
shear viscosity
\beq
\eta \propto T^{-5} \ ,
\label{eta_res}
\eeq
This dependence on temperature is also seen in other systems
where viscosity arises from $2\leftrightarrow 2$ scattering of
Goldstone bosons \cite{Manuel:2004iv,Rupak:2007vp}.
The constant of proportionality in \eqn{eta_res} has mass dimension 8.
In the case of the shear viscosity due to phonons there was only 
one possible scale, the quark chemical potential $\mu_q$,
so $\eta_H \propto \mu_q^8/T^5$.  However, we have several scales
($f_\pi$, $\De$, $m_K$) manifesting themselves in two
coupling constants $\la$ and $g$ \eqn{lambda.g}. Which of these
is most important depends on whether the scattering 
is dominated by the contact term or by the exchange of a virtual particle. 
The dimensionless parameter $u$ that determines which scattering process 
dominates is 
\beq
u =
\frac{3g^2}{\la} = 96\,C\,\sin^2(\varphi) \left(\frac{\mu_K}{\De}\right)^2,
\label{cc}
\eeq
where $\varphi = 2\sqrt{|\dm|/m_K}$ \eqn{phi} and
the $3$ represents the $3$ channels for virtual particle
exchange. For typical values of $\dm$, $m_K$, and $\De$, this ratio can be
bigger or smaller than $1$.  When $u\ll 1$, the contact term
dominates, so the scattering amplitude is proportional to $\la$.  When
$u\gg 1$, the particle-exchange process dominates, so the scattering
amplitude is proportional to $g^2$.
The shear viscosity is inversely proportional to the scattering cross-section,
so
\beq
\ba{rl}
u\ll 1: & \dsp \eta = h_1(\nu)\frac{1}{C^2} \frac{f_\pi^4 \De^4}{T^5} \ ,\\[2ex]
u\gg 1: & \dsp \eta = h_2(\nu)\frac{1}{C^4} 
  \frac{f_\pi^4 \De^8}{\sin^4(\varphi) \mu_K^4 T^5} \ ,
\ea
\label{eta-form}
\eeq
where $h_1$ and $h_2$ are dimensionless functions that depend only
on the Goldstone kaon speed $\nu$, which depends on $\dm$ and $m_K$.  
We can obtain their analytic form when $\nu \ll 1$.  
In that case, the leading-order behavior in both regimes is a $\nu^{11}$
power law (see appendix~\ref{app:coll-int}). In general they must be calculated
numerically, and the result (with the $\nu^{11}$ power law scaled out)
is shown in Fig.~\ref{fig:b12}.

Finally, we note that the shear viscosity due to 
Goldstone kaons will be smaller than that due to the superfluid phonons, since
$f_\pi$, $\De$, and $m_K$ are much less than $\mu_q$.


\subsection{Numerical results}
\label{sec:num.results}

\begin{figure}[htb]
\includegraphics[width=0.75\textwidth]{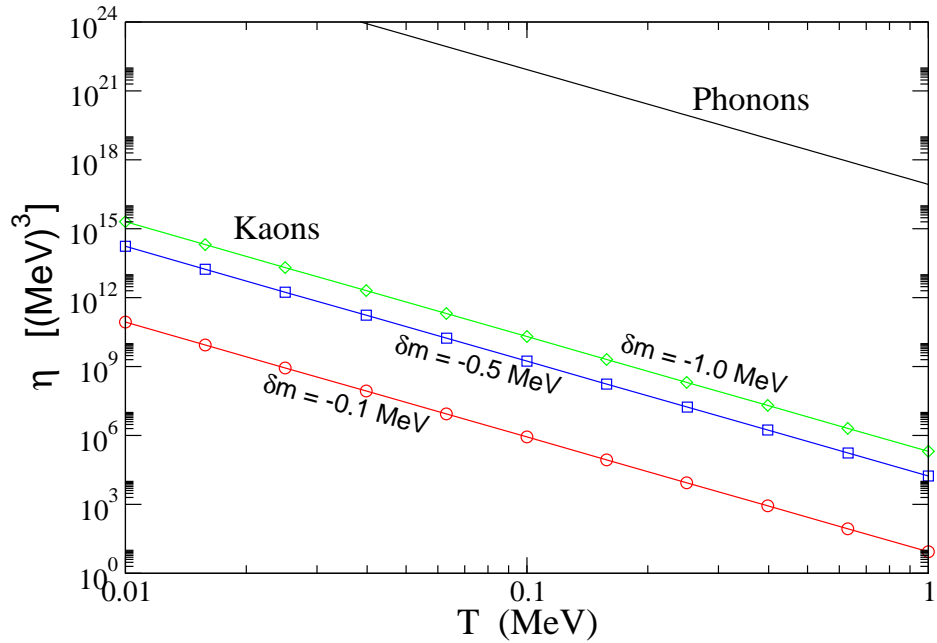}
\caption{
(Color online)
The shear viscosity as a function of temperature for kaons and phonons.
For parameter values, see text.
In the lower part of the graph, the points are numerical calculations and 
the straight lines are fits to the power law form given in \eqn{eta_res}.
The phonons' calculated shear viscosity is many orders of magnitude larger, 
although using the shear mean free path criterion (appendix~\ref{app:mfp}) we
expect them to be non-hydrodynamic in neutron stars at $T\lesssim 1~\MeV$.
\label{fig:shear_T}}
\end{figure}

To begin, we will confirm the temperature dependence predicted
in the previous section.  To do so, we will fix the mass of the kaon,
$m_K = 4~\MeV$, $f_\pi= 100~\MeV$, $\De = 100~\MeV$ and $C=1$.  
In Fig.~\ref{fig:shear_T}, we show the 
viscosity as a function of temperature for a few values of $\de m$.  
The data points are obtained by
numerical evaluation of the 5-dimensional integral, whereas the 
lines show the fit to a $T^{-5}$ power law \eqn{eta_res}, which is
independent of the regime of coupling constants' values.  
On the 
same plot, we show the contribution from the phonons \cite{Manuel:2004iv}. 
As expected, the phonon shear viscosity is much larger.  Most of the difference
comes from the difference in magnitude of the coupling constants (the
kaons coupling constant is larger) and the rest comes from a difference in the 
speed of the kaon and the phonon (the kaon's is smaller), which 
enters the expression for shear viscosity raised to a high power.
Using the shear mean free path criterion $\lshear<1~\km$
(appendix~\ref{app:mfp})
for the validity of hydrodynamics for neutron star oscillations, 
we expect the phonons to be non-hydrodynamic at $T\lesssim 1~\MeV$;
with the parameter values given above,
the Goldstone kaons become non-hydrodynamic at
$T\lesssim 0.03~\MeV$.

\begin{figure}[htb]
\includegraphics[width=0.75\textwidth]{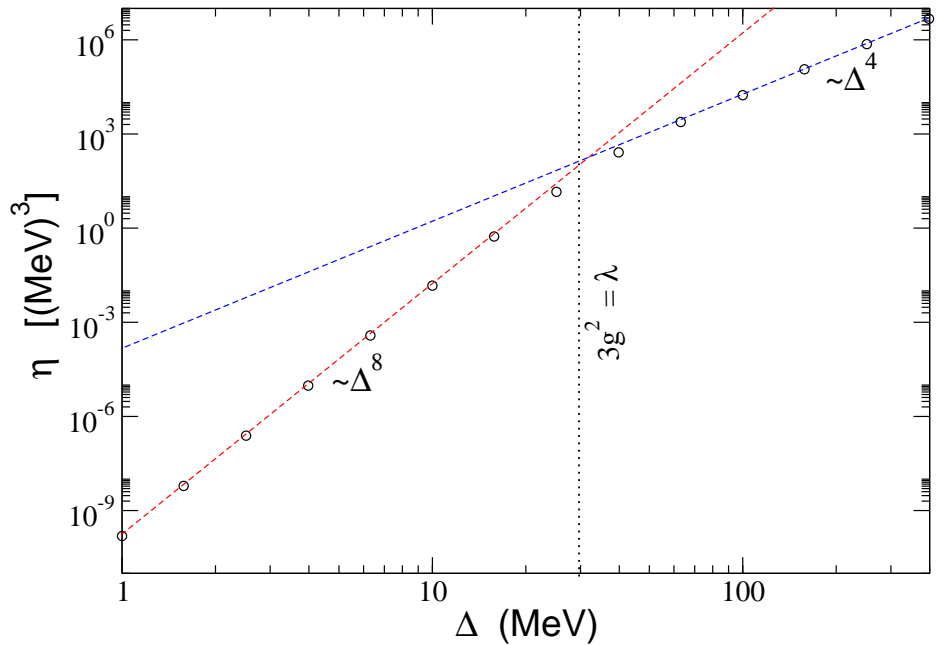}
\caption{
(Color online)
The shear viscosity as a function of $\De$. (See text for
parameter values.)
The points are calculated numerically. The straight lines 
are fits to the power law behaviors of \eqn{eta-form}.
}
\label{fig:shear_D}
\end{figure}

In Fig.~\ref{fig:shear_D}
we show the shear viscosity as a function of the gap $\De$,
with $\de m = -0.5~\MeV$ and $T=1~\MeV$. 
The other parameters have the same values as in Fig.~\ref{fig:shear_T}.
This illustrates the transition between the two regimes given in
\Eqn{eta-form}. The crossover occurs at $u=1$ 
which corresponds to $\De=30~\MeV$, which is indicated on the graph.
As expected, we see that for large $\De$ ($u \ll 1$), $\eta \propto \De^4$;
for small $\De$ ($u \gg 1$), $\eta \propto \De^8$.

\begin{figure}
\includegraphics[width=0.75\textwidth]{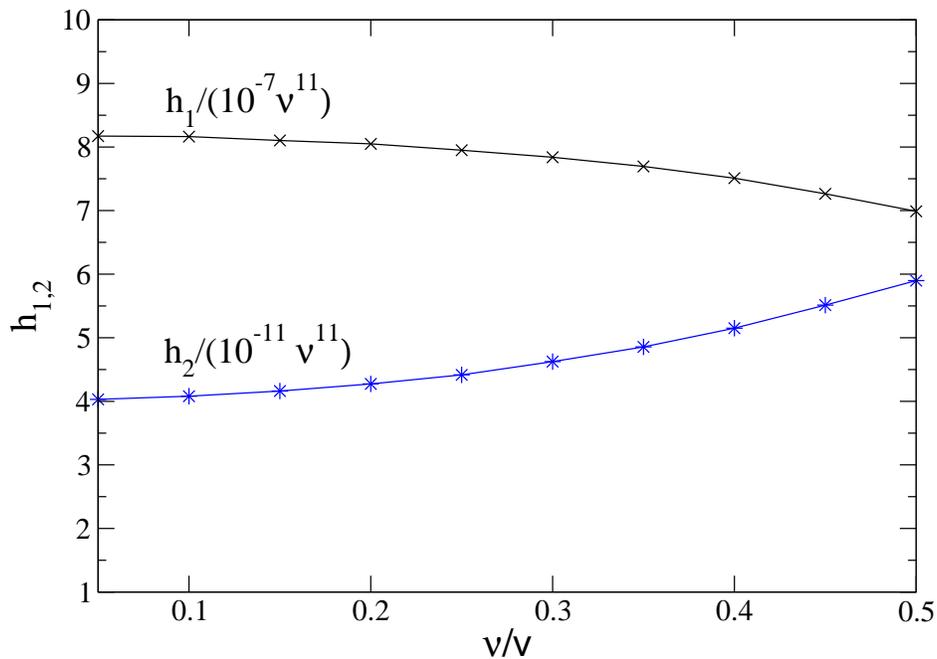}
\caption{
(Color online)
The functions $h_1(\nu)$ and $h_2(\nu)$ \eqn{eta-form}.
We scale out the $\nu^{11}$ power law behavior (see
appendix \ref{app:coll-int}).
On the $x$-axis we show $\nu$ in units of 
the kaon velocity $v\approx \sqrt{|\de m|/m_K}$ 
in the non-kaon-condensed phase. 
}
\label{fig:b12}
\end{figure}

In Fig.~\ref{fig:b12} we present the results of
numerical calculation of $h_{1,2}(\nu)$ \eqn{eta-form}.
We have divided out the dominant behavior $\nu^{11}$ power law
behavior (for details see appendix \ref{app:coll-int}).
We see that the remaining $\nu$ dependence is very mild, so 
to a good approximation the shear viscosity is given by
\eqn{eta-form} with
\beq
\ba{rcl}
h_1(\nu) &\approx& 3.44\ee{-4} \nu^{11} \ ,\\
h_2(\nu) &\approx& 1.70\ee{-8} \nu^{11} \ .
\ea
\label{h1h2}
\eeq

To show  how large the  shear viscosity of CFL-K0 quark
matter could be, we look at a case where the values of the parameters
are pushed in the direction that yields a large value of $\eta$. 
We take $f_\pi,\De \approx 150~\MeV$, $m_K\approx 4~\MeV$, $\dm\sim
-1.0~\MeV$, and $C\approx 0.2$. With these values we find that
at  $T=0.1~\MeV$ ($10^9$\,K)
the shear mean free path \eqn{lshear} is 0.26~km, and 
$\eta = 1.7\ee{13}~\MeV^3 =2.3\ee{18} \,{\rm erg}\,{\rm cm}^{-1}{\rm s}^{-1}$.
At this temperature the phonon's 
shear mean free path (see appendix \ref{app:mfp}) is
larger than the star, so the kaons provide the dominant contribution to
the shear viscosity.

\clearpage

Finally, we check whether the regime of linear hydrodynamics is valid
by evaluating the size of the corrections to the equilibrium stress-energy 
tensor. Linear hydrodynamics is appropriate if $\de T_{ij} \ll T_{ij}$
\eqn{stress}. 
(Note that this is different from the
criterion of validity for hydrodynamics in general, 
discussed in appendix~\ref{app:mfp}.)
This inequality becomes
\beq
\eta \ll V \ell (P+\eps)
\eeq
where $\ell$ represents the length scale of the velocity gradients,
and $V$ is the typical fluid velocity which we assume is of order 1.
If we use the energy density of free quark matter
$\eps \approx 9 \mu_q^4/(4\pi^2)$ which is of order
$10^{10}~\MeV^4$ at $\mu_q\approx 500~\MeV$,
(and $P\lesssim\eps$, which is typically the case), and use the length scale
$\ell\sim 1~\km$ which is appropriate for oscillations of
neutron stars, we find that linear hydrodynamics is valid as long
as
\beq
\eta \ll 10^{25}\,\MeV^3
\eeq
which is easily obeyed by the values of the shear viscosity that
we have calculated for the CFL-K0 phase.


\section{Conclusions}
\label{sec:conclusions}

In this paper, we have calculated the shear viscosity arising from
self-interaction of the Goldstone kaon mode in the CFL-K0
phase of quark matter. The shear viscosity from the
other Goldstone mode, the superfluid phonon, 
has already been explored in Ref.~\cite{Manuel:2004iv}.
We find the same $T^{-5}$ temperature dependence that was found for
the phonons in the CFL phase and for superfluid modes 
in a unitary Fermi gas \cite{Rupak:2007vp}. 
Our final results are the approximate
analytic expressions \eqn{eta-form}, \eqn{h1h2} for the 
shear viscosity due to Goldstone kaons,
and expressions \eqn{lshear} and \eqn{lscat-K}
for the ``shear mean free path'' and ``scattering  mean free path''
of the Goldstone kaons.

Neutron star
oscillations have a length scale in the kilometer range, so 
the phonon and Goldstone kaon fluids in a neutron star
can only be described by hydrodynamics when their mean free paths
are smaller than this.
We argue that the shear mean free path is the appropriate
quantity to use for this purpose (see appendix \ref{app:mfp}).

Because the coupling constants for the 
Goldstone kaons are roughly an order
of magnitude larger than those 
for the superfluid phonon, the shear viscosity and mean free path
of the Goldstone kaon are both several orders of 
magnitude smaller than for the superfluid phonon (see Fig.~\ref{fig:shear_T}).
Using the shear mean free path (see appendix \ref{app:mfp}), 
we find that the superfluid phonons in a neutron star
are described by  hydrodynamics
at temperatures above about 1~MeV ($10^{10}$\,K). 
The Goldstone kaons are hydrodynamic down
to lower temperatures: the exact threshold depends 
sensitively on the
value of the constants in the effective action, but could easily
be lower than the $0.01$ to $0.05~\MeV$ range at which
our treatment becomes invalid because the weak-interaction mass
of the Goldstone kaon \cite{Son:2001xd} must then be taken in to account.

We conclude that, in the temperature range $0.01$~MeV to 1~MeV,
depending on the values of the coupling constants in their effective
theory, Goldstone kaons may very well provide the dominant contribution
to the shear viscosity in CFL-K0 quark matter.

There are several ways in which this work can be developed further.
Firstly, we chose a specific form of the interaction lagrangian \eqn{Lint}
which has one coupling constant, rather than the most general
form \eqn{L-GBint} which has three; our calculation should be
extended to the most general lagrangian. Secondly, it would be useful to
extend our calculation to lower temperatures where, as noted above,
one can no longer neglect the effects of weak interactions on the
dispersion relation of the Goldstone kaon.
It would also be interesting to study shear viscosity from light
kaons in the {\em non}-kaon-condensed CFL phase. These particles
were found to give a large contribution to the bulk viscosity even 
at temperatures as low as a tenth of their energy gap
\cite{Alford:2007rw,Alford:2008pb}.
Thirdly, we neglected scattering between the Goldstone kaons
and the superfluid phonons. It would be interesting to see
if these processes shorten the phonon mean free path and make
a significant contribution to the shear viscosity. 
Fourthly, even though our calculation is open to extension and
improvement in the ways just described, it would be interesting
to perform an analysis along the
lines of Ref.~\cite{Jaikumar:2008kh} to see whether the shear viscosity of the 
Goldstone kaons can have a significant effect on the development
of $r$-modes in a quark star or hybrid neutron star.
Fifthly, as discussed in 
appendix~\ref{app:mfp}, we did not consider how the interactions themselves 
would alter the dispersion relation.  This could affect the calculation
of the mean free path at leading order in the induced non-linearity, but 
would provide a subleading correction to the shear viscosity.
Finally, even when the superfluid phonons or Goldstone kaons 
are not in the hydrodynamic regime, they can still transfer momentum 
over long distances, and it
is important to investigate how they could provide ballistic-regime damping
(as opposed to hydrodynamic viscous damping) of neutron star oscillations.

\section*{Acknowledgements}
We thank Jingyi Chao, Cristina Manuel, Sanjay Reddy, Gautam Rupak,
Thomas Sch\"afer, and Andreas Schmitt
for discussions. This research was
supported in part by the Offices of Nuclear Physics and High
Energy Physics of the U.S.~Department of Energy under contracts
\#DE-FG02-91ER40628,  
\#DE-FG02-05ER41375,  
and \#DE-FG02-03ER41260.  

\newpage


\appendix

\section{Mean Free Path}
\label{app:mfp}

In this appendix we discuss the mean free path of the Goldstone kaon
in the CFL-K$^0$ phase. We expect that hydrodynamics will be applicable 
to neutron star oscillations when
the mean free path is well below the kilometer scale, since neutron star
radii are about 10~km.

We study two definitions of the mean free path, which we call the
``shear mean free path'' $\lshear$ and the 
``scattering mean free path'' $\lscat$. 
The shear mean free path is based on the value of the shear viscosity itself, 
and is probably the physically relevant quantity for deciding when
hydrodynamic calculations of shear viscosity are valid.
The scattering mean free path is the average distance between 
collisions of the Goldstone kaons, including co-linear scattering
events. Since co-linear scatterings do not contribute to the shear viscosity
itself, it seems likely that this quantity is not the relevant one
for finding the limits of validity of  shear viscosity calculations.

\subsection{Shear mean free path}
\label{sec:lshear}

We take our definition of the shear mean free path from 
Ref.~\cite{Rupak:2007vp}, Eq.~(48), where it is referred to as $\la_B$,
\beq
\lshear = \frac{\eta}{n\,\<p\>}
\label{lshear}
\eeq
where $\eta$ is the shear viscosity (given for kaons
by \eqn{eta-form} and \eqn{h1h2}),
$\<p\>$ is the thermal average momentum, and $n$ is the boson density,
\bea
\<p\> &=& 2.7 \, T/\nu \ , \\
n &=& \int \frac{d^3 p}{(2\pi)^3} f_p = \zeta(3)\frac{T^3}{\pi^2\nu^3} \ .
\label{ndef}
\eea
where $\nu$ is the speed of the Goldstone bosons (given for
kaons by \eqn{dispersion}). 

This already allows us to make an estimate of the maximum shear viscosity
that Goldstone kaons can provide, since it follows that 
$\eta \approx 0.3 \nu^{-4} T^4\lshear$, so the maximum shear viscosity
that could possibly occur in a neutron star at temperature $T$
is when $\lshear\approx 1~\km$, i.e.
\beq
\eta_{\rm max} \approx \frac{T^3}{\nu^4} \frac{T}{7\ee{-16}~\MeV} \ .
\eeq
For Goldstone kaons it is quite possible to get $\nu\approx 0.1$
by using small values of $\dm$. Using this
value we find the following upper limits:
at $T=0.01~\MeV$, $\eta_{\rm max}\sim 10^{11}~\MeV^3$;
at $T=0.1~\MeV$, $\eta_{\rm max}\sim 10^{15}~\MeV^3$;
at $T=1~\MeV$, $\eta_{\rm max}\sim 10^{19}~\MeV^3$.

For superfluid phonons we can make more definite statements
because there is less uncertainty about the parameters
appearing in these expressions. The shear viscosity is
$\eta=1.3\ee{-4}\mu_q^8/T^5$ and the speed $\nu$ is generally assumed to 
take its perturbative value $1/\sqrt{3}$ \cite{Manuel:2004iv}, 
in which case we immediately find that
\beq
\lshear_H \approx 4\ee{-5}\frac{\mu_q^8}{T^9} 
\eeq
So phonon hydrodynamics becomes invalid in neutron stars
when $\lshear_H \gtrsim 1~\km$,
i.e. for $\mu\approx 500~\MeV$ we require $T\gtrsim 1~\MeV$ ($10^{10}$\,K),
and at a temperature of 1~\MeV\ the phonon shear viscosity is $5\ee{17}~\MeV^3$.

\subsection{Scattering mean free path}
\label{sec:lscat}

Our calculation of this quantity closely follows that of
Ref.~\cite{Manuel:2004iv}.
The scattering mean free path is determined by the 2-body interaction 
cross section.
The relevant scattering amplitude is the sum of four Feynman diagrams,
the contact term and the s, t, and u-channel diagrams, 
see Fig.~\ref{fig:feyn}. In the s, t, and u-channels there is a
virtual particle which can go on-shell, which means that its self-energy
must be included to avoid an unphysical divergence. 
(The fact that the virtual particle can go on-shell means
that the $2\rightarrow 2$ collision rate already includes
the contribution from $1\rightarrow 2$ splitting process, so these need not
be calculated separately \cite{Arnold:2000dr, Arnold:2003zc}.  We have performed this separate
calculation and verified that the result has the same parametric dependence as the one
we obtain below.)
The rate (per unit volume) for the $2\rightarrow 2$ scattering process is 
\beq
\Ga_{2\rightarrow2} = \half \int_{p,k,p',k'} (2\,\pi)^4 \de^4(P+K-P'-K') 
|\Mscat|^2\,f_p\, f_k\, (1+f_{p'})\, (1+f_{k'}),
\eeq
where
\beq
\int_p = \int \frac{d^3 p}{2\,\eps_p\,(2\,\pi)^3}\qquad \qquad 
f_p = \frac{1}{e^{E_p/T}-1}
\eeq
and 
\beq
|\Mscat|^2 = |{\Mscat_{\rm s}}|^2 + |{\Mscat_{\rm t}}|^2 
+ |{\Mscat_{\rm u}}|^2 + I 
\eeq
where $I$ represents the contact and interference terms, and the matrix elements are 
given in \Eqn{Mscat}. The dominant contribution comes
from $|{\Mscat_{\rm s}}|^2 + |{\Mscat_{\rm t}}|^2 
+ |{\Mscat_{\rm u}}|^2$, since these each have a large enhancement when
the virtual particle is close to being on-shell. The contact term
and interference terms have no such enhancement, and so make
a much smaller contribution to the rate.

We now give a detailed explanation
of the evaluation of $|{\Mscat_{\rm s}}|^2$; the others can be
obtained by similar methods.  We first define the virtual particle momentum 
$Q = (E_q,{\bf q})$ and shift the integral over ${\bf k}$ to an integral over ${\bf q}$. 
We can then still use the momentum-conserving delta-function to do
the integral over ${\bf k'}$.  We then choose the 
direction of ${\bf q}$ as the $z$-axis, so the angular part of the ${\bf q}$ 
integral gives a factor of $4\,\pi$ and the remaining integrand is azimuthally 
symmetric, so the 2 remaining  azimuthal integrals 
give a factor of $2\pi$. This leaves
three integrals over the magnitudes of 
${\bf p}$ , ${\bf k'}$ and ${\bf q}$ as well as over two polar angles, between
${\bf q}$ and ${\bf p}$ and between ${\bf q}$ and ${\bf k'}$.  We then
introduce the auxiliary variable $\om$ via the identity
\beq
\de(E_p + E_k - E_{p'} - E_{k'}) = \int_{-\infty}^{\infty} d\om
\de(\om - E_p + E_{p'})\, \de(\om + E_{k} - E_{k'}).
\eeq
The integral over the two polar angles can then be done using these two 
delta-functions, leaving behind four integrals over $p, k'$, $\om$ and $q$, 
\beq
\Ga_s = A\int_0^\infty dq \int_{\nu\,q}^\infty d\om \int_{(\om - \nu\,q)/2\nu}^{(\om + \nu\,q)/2\nu} dp \int_{(\om - \nu\,q)/2\nu}^{(\om + \nu\,q)/2\nu} dk'\ |{\Mscat}_s|^2\,f_{\nu p}\,f_{\om -\nu p}\,(1+f_{\om -\nu k'})\,(1+f_{\nu k'})
\eeq

To make the temperature dependence explicit we introduce a new set of 
variables ($x,y,w,$ and $z$). For the $s$-channel these are
\beq
x=\frac{\nu p}{T} \qquad \qquad y=\frac{\nu k'}{T}\qquad \qquad 
w=\frac{\om}{T}\qquad \qquad z=\frac{\nu q}{T} \ ,
\eeq 
and the Mandelstam variable is $s = w^2-z^2$.
This leads to the full expression for the s-channel rate, 
\beq
\Ga_s = \frac{g^4 T^{12}}{16 \nu^{17} (2\pi)^5 f_\pi^8} 
  \int_0^\infty \!\!dz \int_z^\infty \!\!dw 
  \int_{(w-z)/2}^{(w+z)/2} \!\!dx \int_{(w-z)/2}^{(w+z)/2} dy
  \, F(x,y,w)\frac{G(x,w,s) G(y,w,s)}{s^2 + (\Pi^+/T^2)^2} 
\label{rate}
\eeq
with
\beq
F(x,y,w) = f_x\,f_{w-x}(1+f_y)(1+f_{w-y}) \qquad
G(x,w,s) = w^2\left(3x(1-\nu^2)(w-x) - s\right)^2
\label{FG}
\eeq 

The self-energy term has both a real and imaginary part, however the real part 
is much smaller~\cite{Manuel:2004iv}.  We will therefore only 
consider the imaginary part ${\rm Im}\ \Pi(\om,q)$, obtained in 
Refs.~\cite{Alford:2007rw,Alford:2008pb,Manuel:2004iv},
\beq
\ba{rcl}
{\rm Im}\ \Pi(w,z) &=&\dsp \Pi^+ \Theta(w^2 - z^2) + \Pi^- 
\Theta(z^2 - w^2)\nn
\Pi^+ &=&\dsp \frac{g^2\,T^6}{16\pi \nu^7 f_\pi^4}\frac{1}{z f_w}\int_{\frac{w-z}{2}}^{\frac{w+z}{2}}dy\ G(y,w,s)\, f_{y}\,f_{w-y}\\[2ex]
\Pi^- &=&\dsp \frac{g^2\,T^6}{8\pi \nu^7 f_\pi^4}\frac{1}{z f_w}\int_{\frac{w+z}{2}}^\infty dy\ G(y,w,t)\, f_{y}\,\left[1+f_{y-w}\right]
\ea
\label{self-energy}
\eeq
where $\Pi^+$ is relevant for the $s$-channel where $w>z$,
and $\Pi^-$ is relevant for the $t$- and $u$-channels where $w<z$.  
We have neglected the tadpole contribution to the self-energy: it
only corrects the kaon velocity by term proportional to $\la T^4$
\cite{Manuel:2004iv}. 

In the remaining 4-dimensional integral \eqn{rate}
we can now use a simple approximation to greatly simplify the integral.  
In the expression for $\Ga_s$ the integral over $w$
is sharply peaked at the limit of integration
where $w=z$, i.e.~$\om = \pm \nu q$ ($s=0$).
The integral takes the form 
\beq
\int_z^\infty dw\frac{I(w,z)}{(w^2-z^2)^2 + (\Pi^+(w,z)/T^2)^2} \approx
  \frac{\pi T^2}{4\, \Pi^+(z,z)\, z} I(z,z) \ ,
\label{approx}
\eeq
This expression is valid when $I(w, z)$ is slowly varying near the 
singular point $w=z$ and when $\Pi^+/T^2 \ll 1$ because $\Pi \sim g^2 T^6$
and $T \ll f_\pi, \De, \mu_K$.
Applying this approximation to the $s$-channel contribution, we find
\beq
\Ga_s = \frac{g^4 T^{14}}{16 \nu^{17} (2\pi)^5 f_\pi^8}\frac{\pi}{4}\int_0^\infty dz \int_0^z dx \int_0^z dy f_x f_{z-x} 
(1+f_y) (1+f_{z-y}) \frac{G(x,z,0) G(y,z,0)}{\Pi^+(z,z) z}.
\label{ga_s}
\eeq
where
\beq
\Pi^+(z,z) = \frac{g^2 T^6}{16 \pi \nu^7 f_\pi^4}\frac{1}{z f_z} \int_0^z dx\ f_x\, f_{z-x}\, G(x,z,0)\ .
\eeq
We can then see that the integral over $x$ in \Eqn{ga_s} partially cancels the
integral contained in $\Pi^+$ leaving behind only a double integral. Using
$G(y,z,0)$ from \Eqn{FG}, we have
\beq
\Ga_s = \frac{g^2 T^8 (1-\nu^2)^2}{\nu^{10}f_\pi^4} J\ ,
\eeq
where $J$ is a pure number given by
\bea
J &=& \frac{9}{128 \pi^3} \int_0^\infty dz\ \int_0^z dy\ y^2\,z^2\,(z-y)^2\, f_z\,(1+f_y)\,(1+f_{z-y}) \nn
 &\approx& 0.466 \ .
\eea
Applying the same method to the $t$- and $u$-channel integrals, 
we find that they all give the same rate, so the total rate is
\beq
\Ga_{\rm total} = 1.40\, \frac{g^2\, T^8\, (1-\nu ^2)^2}{\nu ^{10}f_\pi^4} \ .
\eeq
The scattering mean free path is defined as
\beq
\lscat_K = \frac{\nu n}{\Ga} \ ,
\eeq
where $n$ is the particle density \eqn{ndef}.
So the scattering mean free path of the Goldstone kaons is
\beq
\lscat_K = 0.0881 \frac{\nu^8}{(1-\nu^2)^2} \frac{f_\pi^4}{g^2 T^5}
= 3.44\ee{-4} \frac{\nu^8}{(1-\nu^2)^2} \frac{f_\pi^2\,\De^4}{C^2\,\mu_K^2\,\sin^2{\varphi}}
T^{-5} \ ,
\label{lscat-K}
\eeq

For comparison, the scattering
mean free path of the phonon is~\cite{Manuel:2004iv}
\beq
\lscat_H = 0.181\, \frac{v^8}{(1-v^2)^2} \frac{\mu_q^4}{T^5} 
  = 5.02\ee{-3} \frac{\mu_q^4}{T^5}
\label{lscat-H}
\eeq
so the ratio of the two is
\beq
\frac{\lscat_K}{\lscat_H} = 
 17.546 \frac{\nu^8}{(1-\nu^2)^2} \frac{f_\pi^4}{g^2 \mu_q^4} 
\approx  2.14\ee{-4} \frac{f_\pi^2 \De^4 |\de m|^3}{C^2 m_K^5 \mu_q^4} \ .
\eeq
Since $\mu_q$ is much larger than any of the other energy scales,
this implies that the scattering mean free path of the kaon,
like the shear mean free path, is generally 
much shorter than that of the phonon, giving the Goldstone kaon
a much wider range of temperatures where it can be treated hydrodynamically.

We have noted above that we expect the shear mean free path to be a
better indicator of the range of validity of hydrodynamics, but for the sake
of completeness we now estimate the temperature at which $\lscat_K$
will become greater than 1~km, in the case of very unfavorable
parameter choices that lead to a long mean free path.
We will use the values used at the end of Sec.~\ref{sec:num.results}
to illustrate how high the shear viscosity can be, namely
$f_\pi=150~\MeV$, $\De=150~\MeV$, $\dm=-1.0~\MeV$, $m_K=4.0~\MeV$ 
and $C = 0.2$.  In this case the scattering mean free path is shorter than 1~km 
for $T\gtrsim 0.006~\MeV$. For more favorable choices of the
couplings this critical temperature will be much lower.
In comparison, from \eqn{lscat-H} the scattering
mean free path for phonons is shorter than 1~km for $T> 0.04~\MeV$.

We have used a linear dispersion for the kaon in calculating the mean
free path, which is a requirement for getting a co-linear enhancement.  
However, there are sources of non-linearities in the dispersion.  One
comes directly from our expansion of the full kaon dispersion in 
\eqn{dispersion}.  If we had kept higher order terms, we would
get a contribution that behaves as 
\beq
E = \nu p (1 + \ga p^2)
\eeq
where $\ga>0$.  This positive curvature would still allow for the
co-linear splitting and joining processes.  Therefore, keeping this
term would provide a subleading contribution to the calculation
presented here.  

However, we ignored how the higher order derivative interactions
themselves could change the kaon dispersion.  Something similar
has been calculated for the superfluid phonons, \cite{Zarembo:2000pj},
where $\ga$ was found to be negative and therefore the $1\leftrightarrow2$
processes are kinematically forbidden.  If the corresponding
non-linearity for the kaons were positive, then as above, the calculation
presented here would remain the same.  However, if the curvature
were negative as for the phonons, then the mean free path would be altered
at leading order.  This is basically because the non-linearity itself
would act to regulate the on-shell propagator and the scattering rate
would go like $1/\ga$ instead of $1/\Pi$ (where $\Pi$ is the self-energy).
The appropriate scales to compare are $\ga T^2$ and $\Pi/T^2$ and in the
case of the phonons, $\ga T^2 \gg \Pi/T^2$, such that this correction would 
make the mean free path even larger and affect the validity of hydrodynamics.  
See \cite{Braby:2009dw} for a calculation involving the non-linear phonon 
dispersion and its affect on regulating the phonon propagator
in a calculation of the thermal conductivity.  However, it should
be noted that including the non-linearity would only provide
a subleading correction to the shear viscosity for either sign of $\ga$
because the shear viscosity is insensitive to that region of phase space.


\section{Power counting sharply peaked integrals}
\label{app:integrals}

Here we discuss in more detail the evaluation of integrals of the
type \eqn{approx}, having
a slowly-varying component $I$ multiplied by a function with a sharp
Lorentzian peak at the edge of the range of integration.
In appendix~\ref{app:mfp} we assumed that $I$ was non-zero at
the edge of the range and we kept only the leading contribution.
Here we include higher-order corrections by
Taylor-expanding the numerator,
\beq
\int_0^z dw \frac{I(w,z)}{(w^2-z^2)^2 + \eps^2} \sim \int_0^z 
\frac{I(z,z) + (w-z)I'(z,z) + \half(w-z)^2 I''(z,z) + \dots}
{(w^2-z^2)^2 + \eps^2} \ ,
\label{approx2}
\eeq
where $I'(z,z)$ is the first derivative of $I$ with respect to $w$, evaluated 
at $w=z$.  We then find
\bea
J_1 \equiv \int_0^z dw \frac{1}{(w^2-z^2)^2+\eps^2} &\sim& \frac{\pi}{4z\eps}\nn
J_2 \equiv \int_0^z dw \frac{z-w}{(w^2-z^2)^2+\eps^2} &\sim& 
             -\frac{{\rm ln} \eps}{4z^2} \nn
J_3 \equiv \int_0^z dw \frac{(z-w)^2}{(w^2-z^2)^2+\eps^2} &\sim& \frac{1}{8z}
\label{full_asymp}
\eea
This gives us a scheme for power counting any integrals of the form given by
\eqn{approx2}. The relevant property is the dependence on $\eps$, since
the collision integrals for transport properties take the form
\eqn{approx2} with $\eps = \Pi/T^2 \propto g^2 T^4 \ll 1$.  

We can now justify the statement made in Sec.~\ref{sec:shear} that when we 
calculate the shear viscosity using a polynomial expansion of the
function $g(p)$ \eqn{g-pnml}, the dominant contribution comes from $g(p)=1/p$,
i.e.~choosing the minimum-exponent parameter $n$ to be $-1$.

Calculations of the mean free path
and the shear viscosity both involve a rate calculation
which contains collision integrals.
In the mean free path collision integral  \eqn{rate}
there is a sharp peak in the integrand at $w=z$ 
corresponding to a co-linear divergence, 
where two kaons have parallel momenta, and exchange
a kaon whose momentum lies in the same direction.
In the mean free path calculation this near-divergence is regulated by the
self-energy, so the result depends on the self-energy ($\sim 1/\eps$).

In the case of the shear viscosity, we expect the integral not
to have a co-linear divergence, since shear viscosity measures
momentum transfer, so processes that do not change the momentum direction
of the particles make no contribution. We therefore expect that
the collision integrand in 
the shear viscosity should go to zero at $w=z$ in such a way that
the result does not depend on the self-energy.
The true physical $g(p)$ function will give an integrand
that has this property. However, if we make a bad guess at
$g(p)$ (by using inappropriate basis polynomials in the
expansion \eqn{g-pnml}) then each individual term will have
a co-linear divergence, which will only cancel out when we
add up the contributions from many terms.
The best guesses for $g(p)$ are therefore
ones which yield a collision integrand with
no co-linear divergence, i.e~no dependence on the self-energy.

In both the mean free path and the shear viscosity calculations
the collision integral takes the form \eqn{approx2}. The difference
between them lies in behavior of the numerator $I(w,z)$ in the
co-linear regime $w\to z$. In the mean free path calculation
(appendix \ref{app:mfp}), the numerator stays finite in this regime,
so the collision integral is of the form $J_1$ \eqn{full_asymp}
and is strongly dependent on the self-energy.
In the shear viscosity calculation, the numerator contains 
an additional factor $\De^s_{ij} \De^t_{ij}$ (see \Eqn{Mst}), 
which we'll call the transport term. The behavior of the
transport term in the co-linear regime is therefore crucial
in suppressing the self-energy dependence.

Using expansions of $g(p)$ with only one term ($N=0$) we calculate
$\De^0_{ij}$ and the shear-viscosity
collision integral for different choices of $n$.
We summarize the results in Table~\ref{tab:colinear_suppression}.
We see that the choice $n=-1$ fully suppresses
the co-linear singularity and gives a collision
integral that is independent of the self-energy.  The choice
$n=-2$ partially suppresses co-linear scattering and gives
the collision integral a very weak dependence on the self-energy.
Other values of $n$ do not
suppress the co-linear scattering at all and are akin to the calculation
of the mean free path.  
This explains our finding in Sec.~\ref{sec:shear}, Table~\ref{tab:res} 
that $n=-1$ is the the optimal choice for fast convergence of the
polynomial approximation for $g(p)$ in the shear viscosity, that
$n=-2$ is the next best choice, and other values of $n$ have very
poor convergence.

\begin{table}[tb]
\begin{tabular}{ccc}
\hline
\rule[-2.5ex]{0em}{6.5ex} 
\parbox{10em}{Min-exponent\\[-2ex] parameter $n$} 
 & \parbox{10em}{Behavior of \\[-2ex] $\De^0_{ij}$ near $w=z$}  
 & \parbox{10em}{$\eps$-dependence of \\[-2ex] collision integral}\\
\hline
$n = -1$ & $(z-w)^2$ & independent \\
$n = -2$ & $(z-w)$ & $-\ln \eps$ \\ 
$n\neq -2,-1$ &  $(z-w)^0$ & $1/\eps$  \\ 
\hline
\end{tabular}
\caption{Table of behavior of $\De_{ij}$ (part of the collision integrand)
near the co-linear singularity, and the collision integral; $\eps$
represents the self-energy. Only the $n=-1$ case has the proper physical
suppression of co-linear contributions to the shear viscosity.
}
\label{tab:colinear_suppression}
\end{table}

\section{Approximate evaluation of the collision integral}
\label{app:coll-int}

Here we describe how the collision integral is reduced to a
five-dimensional numerical integral, in which we have
factored out the temperature dependence and
part of the dependence on the kaon speed $\nu$.  
We begin with the matrix $M$ \eqn{Mst}
that enters in to the calculation of the shear viscosity.
As described in Sec.~\ref{sec:shear} and 
appendices~\ref{app:mfp} and \ref{app:integrals}, 
we get a good estimate
of the shear viscosity by assuming $g(p)=1/p$, i.e.~we
set $N=0$ and $n=-1$. Then, as described after \Eqn{prop}, we can 
eliminate seven of these integrals by using the $\de$-function 
and spherical symmetry. We rescale the momenta with temperature, and find
\beq
M_{00} = \frac{1}{10\cdot 2^8\,\pi^6\,\nu^6}\left(\frac{T}{\nu}\right)^{13}
\int d\Ga \ f_x\,f_y\,(1+f_z)\,(1+f_w) |\overline{\Mscat}(\nu,g,\la)|^2 \,
\bar{\De}^0_{ij}\, \bar{\De}^0_{ij}
\label{M00-rescaled}
\eeq
where  $f_x\equiv 1/(e^x-1)$, and
\beq
\int d\Ga = \int_0^\infty dx \int_0^\infty dy\, \int_{-1}^{1} d\al \int_{-1}^{1} d\be
\int_0^\pi d\phi\ \frac{z^2}{1-\al}\ ,
\eeq
\beq
x = \frac{\nu p}{T} \qquad y = \frac{\nu k}{T} \qquad z = \frac{\nu k'}{T} = \frac{x\,y\,(1-\al)}{x(1-\be) + y(1-\ga)}\qquad w = \frac{\nu p'}{T} = x+y-z\ ,
\eeq
\beq
\al = {\bf \hat{p}} \cdot {\bf \hat{k}} \qquad \be 
  = {\bf \hat{p}} \cdot {\bf \hat{k'}} 
\qquad \ga = {\bf \hat{k}} \cdot {\bf \hat{k'}} = 
\al\be + \sqrt{(1-\al^2)(1-\be^2)}\cos(\phi)
\eeq
From \eqn{Mscat}, $\overline{\Mscat}\equiv (\nu/T)^8 \Mscat$ 
depends on the speed 
$\nu$ and the couplings $g$ and $\la$ as well as the rescaled momenta 
$x$, $y$, $z$. From \eqn{Delta_ij}, 
$\bar{\De}^0_{ij} \equiv (\nu/T)\, \De^0_{ij}$ 
depends only on the rescaled momenta.
The expression for $z$  comes from
solving the energy-conserving $\de$-function, and $\phi$ is the
difference in azimuthal angles between the vectors ${\bf p}, {\bf k}$
and ${\bf p}, {\bf k'}$.


We have scaled out all the temperature dependence of the integrand,
but there is still some dependence on $\nu$ and the couplings $g$ and $\la$
which comes in via $\Mscat$.
The integral can be evaluated numerically using \eqn{Delta_ij}
and \eqn{Mscat} for given values of $\nu$, $g$, $\la$.

We now show how to obtain the approximate analytic forms
for $\eta$ given in \eqn{eta-form},
which are valid in the 
regime $g^2/\la\ll 1$ and $g^2/\la\gg 1$.

Using \eqn{M00-rescaled} in \eqn{eta.full} and evaluating
$A^{-1}_0$ from \eqn{An},
we find that
\beq
\eta = \frac{\rm const}{\nu^8\,M_{00}(\nu)}
\eeq
where const represents a function independent of $\nu$.
We can obtain the function $h_1(\nu)$ in \eqn{eta-form} by 
going to large $\la$ in which case ${\Mscat} = {\Mscat}_c$, and
then calculating the shear viscosity with all the dimensionful
parameters in the coupling constant $\la$ set equal to unity.
To calculate $h_2(\nu)$, we go to large $g$ where 
 ${\Mscat} = {\Mscat}_s + {\Mscat}_t + {\Mscat}_u$ and do the
same thing. In both cases we find
\beq
h_{1,2} = \frac{A_{1,2}\,\nu^{11}}{\int d\Ga f(\Ga) |{\Mscat}_{1,2}(\nu)|^2}
\eeq
where $A_{1,2}$ is a pure number and $f(\Ga)$ 
represents the parts of the integrand of \eqn{M00-rescaled} that
are independent of $\nu$. From \eqn{Mst} we find that
\beq
{\Mscat}_{1,2} \sim c^{(0)}_{1,2} + c^{(1)}_{1,2}\nu^2 + c^{(2)}_{1,2}\nu^4
\eeq
Therefore, we expect 
\beq
h_{1,2}(\nu) = C_{1,2}\frac{\nu^{11}}{\sum_{i=0}^4 a^{(i)}_{1,2} \nu^{2i}}\ ,
\eeq
justifying our statement in the paragraph below \Eqn{eta-form} and the 
resulting scaling in Fig.~\ref{fig:b12}. 

Finally, we can explain why $h_1$ and $h_2$ are so small 
(see Fig.\ref{fig:b12}).
This is a direct result of $M_{00}$ being large. Because the
all interactions of the Goldstone kaons  are derivative interactions,
the collision integral involves high powers of momenta.
Schematically, it has the form 
\beq
\int_0^\infty dx\ x^d f_x \sim (d+1)!
\eeq
where $d=12$ in our case. Since $1/13!\approx 1.6\ee{-10}$ it is not surprising
that $h_1$ and $h_2$ are of that order.

\renewcommand{\href}[2]{#2}

\bibliographystyle{JHEP_MGA}
\bibliography{shear,non_spires}

\end{document}